%% file: main.tex
\documentclass[runningheads]{llncs}

\usepackage{eccv}

\usepackage{eccvabbrv}

\usepackage{graphicx}

\usepackage[accsupp]{axessibility}  

\usepackage[pagebackref,breaklinks,colorlinks,citecolor=eccvblue]{hyperref}

\usepackage{orcidlink}

\input{preamble}

\newcommand{\methodname}{SeidelConv\xspace}

\begin{document}

\title{Broadband Wide Field of View Imaging with Computational Mirrors} 

\titlerunning{computational mirrors}

\author{Vishwanath Saragadam\inst{1} \and
Niki Nezakati\inst{1} \and
Amit Roy-Chowdhury\inst{1} \and Vivek Boominathan\inst{2}}

\authorrunning{V.~Saragadam et al.}

\institute{University of California Riverside \and
Rice University\\
\email{vishwans@ucr.edu}}

\maketitle

\begin{abstract}

\input{abstract}
\end{abstract}

\section{Introduction}
\label{sec:intro}
\input{intro}

\section{Prior Work}
\label{sec:prior}
\input{prior}

\section{Computational Mirrors}
\label{sec:method}
\input{method}

\section{Results}
\label{sec:results}
\input{results}

\section{Discussions and Conclusion}
\label{sec:conclusion}
\input{conclusion}

\section{Acknowledgement}
\label{sec:ack}
This work is supported in part by DTRA grant HDTRA12610007, CNS-2312395, CMMI-2326309, and a University of California Regents Faculty Fellowship.

\appendix

\section{Hardware Details}
\label{sec:prior}
\input{supp/hardware}

\section{Results}
\label{sec:results}
\input{supp/results}


%
%
\bibliographystyle{unsrt}
\bibliography{main}
\end{document}

%% file: preamble.tex
\usepackage{soul}

\usepackage[font=small,labelfont=bf]{caption}
\usepackage{booktabs}
\usepackage{multirow}
\usepackage{tabularx}
\usepackage[font=small,labelfont=bf]{subcaption}
\usepackage{comment}

\usepackage{xcolor}

\newcommand{\bfp}[0]{\mathbf{p}}

%% file: abstract.tex
Traditional glass-based optics are typically optimized for narrow spectral bands, such as the visible (400–700nm) or shortwave infrared (1000–1800nm). While the emergence of VIS-SWIR sensors (400–1700nm) offers transformative potential, refractive optics struggle to focus this entire range simultaneously. Mirrors represent a promising achromatic alternative; however, they are often sidelined by field curvature, and off-axis aberrations.
This paper introduces \textbf{Computational Mirrors}, a framework that enables high-resolution, wide-field-of-view imaging across the complete VIS-SWIR spectrum using a single sensor. Our method is built on the observation that distinct regions of the field of view reach focus at varying distances from the mirror. By capturing a minimal focal stack (2–4 images), we utilize a computational backend to recover a sharp, all-in-focus image.
A key contribution of this work is \textbf{\methodname}, a novel, physics-inspired, spatially-varying point spread function (PSF) model designed to accurately characterize and correct the off-axis aberrations inherent in simple concave mirrors. We demonstrate the efficacy of our approach using a first-of-its-kind 50mm F/1 optical system equipped with a VIS-SWIR sensor. Our system produces sharp images across RGB, NIR, and SWIR wavelengths without requiring refocusing, revealing material details invisible within individual spectral bands. We further validate the scalability of our approach with a 100mm F/2 system optimized for long-range imaging.

%% file: intro.tex
\begin{figure}[!tt]
    \centering
    \includegraphics[width=0.95\columnwidth]{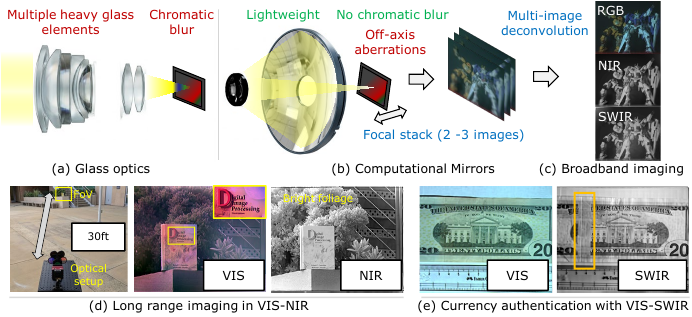}
    \caption{\textbf{Computational mirrors.} (a) Traditional refractive (glass) optics require complex, multi-element assemblies that are inherently bulky and prone to chromatic aberration across the VIS-SWIR spectrum. In contrast, reflective optics are achromatic but suffer from severe off-axis geometric aberrations. (b) Our approach captures a sparse focal stack using simple mirror optics and employs a multi-image deconvolution backend to recover a sharp, wide-field-of-view image. (c) The proposed system enables high-performance, ultra-broadband imaging (400–1700nm) using a compact hardware footprint. (d) Long-Range Imaging: Demonstration of sharp, multi-spectral capture at significant distances across visible and NIR bands. (e) Material Analysis: An application in currency authentication, where imaging across visible and SWIR wavelengths reveals security features invisible to standard sensors.}
    \label{fig:teaser}
\end{figure}

Recent advancements in sensor technology have unlocked imaging capabilities across an expansive spectral range, spanning from the visible to the short-wave infrared (SWIR) wavelengths. These broadband sensors enable high-accuracy material classification~\cite{bioucas2013hyperspectral, paoletti2019deep, bhargava2024hyperspectral}, robust imaging in extreme weather conditions due to low scattering in SWIR~\cite{junwei2013study}, and increased visibility in the dark~\cite{krieg2019comparative}. 
However, the optics required to focus such broadband light remain a bottleneck. Traditional refractive (glass-based) optics are inherently limited by chromatic aberration, where different wavelengths focus at different depths due to the dispersive nature of glass (Fig.~\ref{fig:teaser}(a)).
To mitigate this, optical designers rely on complex, heavy, and often prohibitively expensive multi-element assemblies to align the broad spectral range onto a single focal plane. This complexity frequently leads to undesirable trade-offs, such as restricting the system to a narrow subset of wavelengths or bulky multi-camera architectures to cover the full spectrum.

Reflective optics, including concave mirrors and complex multi-mirror objectives~\cite{korsch2012reflective}, are inherently achromatic from the deep UV to the far-infrared. This characteristic makes them theoretically ideal for VIS-SWIR imaging. Consequently, reflective designs are standard in specialized fields such as microscopy~\cite{norris1951reflecting} and large-scale astronomical instrumentation, most notably the James Webb Space Telescope (JWST)~\cite{gardner2023james}.
However, mirrors, particularly simple configurations with only one or two focusing elements, suffer from severe off-axis aberrations that constrain their versatility. These distortions typically limit the usable field of view (often to 1mm or less in microscope objectives) or necessitate high f-numbers ($f/6$ or higher), which significantly reduces light throughput.
One particularly debilitating effect in few-mirror systems is the Petzval (field) curvature~\cite{kingslake2009lens}. In these systems, the image surface is non-planar; different regions across the field of view reach focus at varying distances from the mirror. This characteristic renders traditional mirror-based solutions impractical for medium focal lengths (50–100mm), where a flat sensor cannot simultaneously capture a sharp image across the entire frame.



This paper presents \textbf{Computational Mirrors}, a practical imaging pipeline that combines hardware-controlled focal stack acquisition with a physics-motivated forward model and multi-image reconstruction. Our approach is based on two primary contributions: First, we demonstrate that while a single exposure from a simple mirror optic suffers from severe field curvature, a focal stack effectively captures sharp regions of the scene at varying axial positions. By capturing a minimal stack (2–4 images), we ensure that every point in the wide field of view is sampled at or near its focus (Fig.~\ref{fig:teaser}(b)). We then employ multi-image deconvolution to fuse these measurements into a single, high-fidelity, all-in-focus image.
%
%
However, these raw focal stack images are still degraded by complex off-axis aberrations. Traditional restoration methods, such as patch-wise convolution or Eigen-decomposed PSF models, are insufficient for modeling these spatial variations inherent in high-aperture mirror systems.
To address this, we propose \textbf{\methodname}, a novel physics-inspired PSF model. Drawing inspiration from Spatial Transformer Networks~\cite{jaderberg2015spatial}, \methodname accurately characterizes the spatially-varying blur across each image in the stack.
\methodname is built on the observation that off-axis aberrations can be effectively modeled as localized geometric distortions. The input image is warped by a small set of affine transformations (typically 20 - 30) that models the distortions inherent to off-axis aberrations. These images are then blurred by learned kernels, and then weighed by per-pixel weights and summed.
These parameters, including the affine matrices, blur kernels, and weight maps, are optimized using a monitor-based calibration procedure~\cite{lin2025learning}. Finally, we integrate the \methodname forward model into a Plug-and-Play (PnP) framework~\cite{zhang2021plug} to recover the final sharp image, leveraging deep priors for robust regularization (Fig.~\ref{fig:teaser}(c)).

We validate the efficacy of Computational Mirrors through experimental prototypes using $f/1$ 50mm and $f/2$ 100mm concave mirror objectives. These are paired with a Sony IMX990 VIS-SWIR sensor, capable of sensing from 400nm to 1700nm. Across a diverse array of real-world scenes, ranging from macro-scale material analysis of currency to long-range outdoor environments, our system captures high-fidelity imagery across the Visible (400–700nm), NIR (700–1000nm), and SWIR (1000–1700nm) bands (Fig.~\ref{fig:teaser}(d,e)). Our approach eliminates the need for wavelength-dependent refocusing, a common failure point in refractive broadband systems. These results represent the first demonstration of high-aperture ($f/1$), ultra-broadband imaging using simple reflective optics. By overcoming the traditional limits of field curvature and off-axis aberrations, this work paves the way for a new generation of compact, lightweight, and cost-effective solutions for multi- and hyperspectral imaging.
We will release the code of \methodname and extensive validation data for further research.



%% file: prior.tex
Our work builds on results in aberration theory of curved surfaces such as mirrors and lenses, focal stacking and computational imaging, and broadband imaging. We explore the prior work in these areas next.

\begin{figure}[!tt]
    \centering
    \includegraphics[width=\columnwidth]{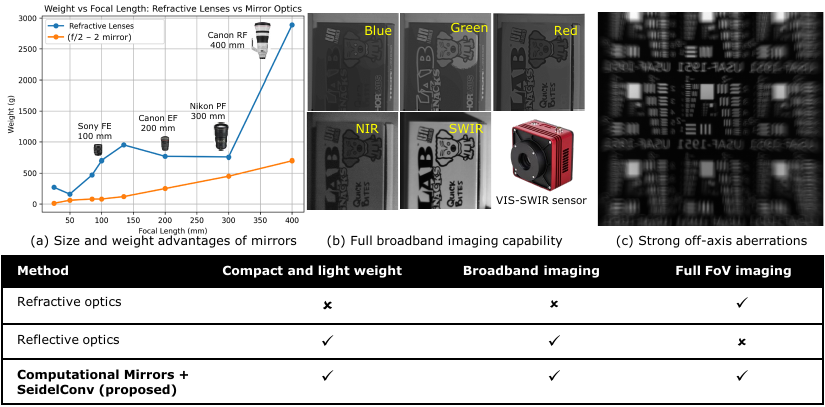}
    \caption{\textbf{Challenges and opportunities with mirrors.} Mirrors have two unique advantages. (a) they are considerably lighter in weight and size compared to refractive optics, and (b) they can focus light across extremely broadband wavelengths. When equipped with VIS-SWIR sensors, they can enable broadband imaging with no wavelength-dependent corrections. However (c) mirrors suffer from strong off-axis aberrations where only the center is in focus. We present a computational approach to obtain sharp images from such aberrated images using mirror-based optics.}
    \label{fig:size_plot}
\end{figure}

\paragraph{Optical aberrations.} Optical aberrations are broadly categorized under geometric and chromatic aberrations. Geometric aberrations occur due to light rays of a single wavelength not focusing at a single point either on axis or off-axis. These aberrations are categorized by Seidel aberrations~\cite{BornWolf1999}.
Additionally, these optics may suffer from the Petzval field curvature, where the focused surface is not planar but a curved one.
These aberrations are common to both glass and mirror optics, and are often corrected with multi element design~\cite{korsch1977anastigmatic}.
%
Chromatic aberrations occur when light of various wavelengths do not focus at the same point, resulting in longitudinal (focal shift), and sometimes off-axis aberrations.
This is inherent to glass-based optics, and is corrected by using glass optics with various refractive index to cancel out the aberrations.
%
%
Invariably, while glass-based optics can produce very large field of view images, they are limited to a narrow range of wavelengths.
Mirrors have no chromatic aberrations and are compact, but suffer from off-axis aberrations and field curvature (see Fig.~\ref{fig:size_plot}).

\paragraph{Mirror-based imaging optics.}
Mirror-based objectives are the workhorse of astronomy, where broadband focusing, high light throughput, and light weight are critical. 
To this end, most observatories, and space telescopes (Hubble, Roman, JWST) are equipped with mirror-based optics, often with 3 or more elements in a so-called anastigmat configuration to obtain diffraction limited performance across the whole field of view. However, these telescopes are often operated at f/10--f/20 and thus have very long focal lengths to image at high spatial resolution.
In the microscopic world, commercial off-the-shelf reflective objectives enable achromatic imaging from 400nm - 20um. These objectives are in the Schwarzchild configuration~\cite{norris1951reflecting} which consists of a concave and a convex mirror with the same curvature to achieve sharp images over the field of view.
However, these objectives have extremely narrow field of view due to strong Petzval field curvature.
%
%
These limitations often lead to catadioptric~\cite{bahrami2010all} solutions where mirrors are combined with lenses to get large field of view~\cite{villa1972adaptation}. 
However, the presence of a lens causes chromatic aberrations, invariably leading to blurry results.
It is possible to build curved sensors that approximately follow the Petzval curvature of mirrors, but only works for a fixed mirror objective. 
%
%
To this end, for imaging macro world, and for hyperspectral objectives, there are no known purely mirror-based objectives between 50-100mm with large light collection capabilities that work with flat sensors.


\paragraph{Computational corrections for imaging with mirrors.} 
Yokoya and Nayar~\cite{yokoya2015extended} demonstrated that a focal stack of images, when summed up, enabled spatially invariant blur, which can then be input to a deconvolution algorithm. In lens-based optics, Matsunaga and Nayar~\cite{matsunaga2015field} showed that a dense focal stack can be leveraged to evaluate an image on the Petzval curved surface, and then can be deconvolved with a spatially invariant blur. Other work achieves snapshot multi-focus capture through beam splitters or specialized optics \cite{abrahamsson2013fast}, trading light efficiency for acquisition speed.
Most prior works have relied on dense focal stacks (10 - 30 images) to enable a sharp image, which is often impractical in real settings.
We build on these ideas but tailor our method to the structure of mirror-based imaging.
Rather than collapsing to a single image (by summing or evaluating on the petzval surface), we perform a multi-image deconvolution with the focal stack. In such conditions, we show that as few as 3 images suffice to obtain a very sharp image, even with F/1 optics, enabling high throughput, broadband imaging capabilities.

%% file: method.tex
We now detail the hardware and algorithm used for overcoming the physical limitations of single element mirror-based objectives.

\subsection{Construction of Mirror Objective}
\begin{figure}[!tt]
    \centering
    \includegraphics[width=\columnwidth]{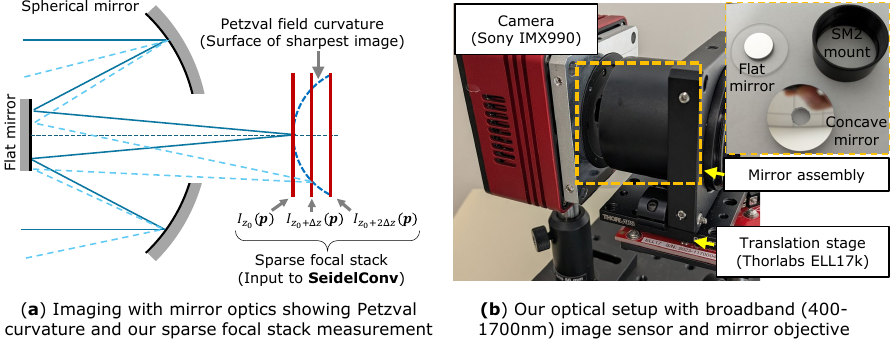}
    \caption{\textbf{Computational mirrors hardware.} (a) Mirror optics enable imaging without any chromatic aberration but suffer from severe field curvature. To tackle this, we leverage a computational reconstruction approach by measuring a sparse focal stack that spans the whole Petzval curved field, and then perform a multi-image deconvolution with our proposed \methodname. (b) shows our optical setup with a broadband VIS-SWIR sensor (ToupTek SensWIR 1.3MP), a translation stage to capture the focal stack, our constructed mirror objective, and five spectral filters spanning blue, green, red, NIR, and SWIR to demonstrate broadband imaging capabilities.}
    \label{fig:construction}
\end{figure}
Our objective is engineered to mimic the form factor and optical alignment of traditional refractive lenses, ensuring seamless integration as a drop-in replacement within standard imaging architectures (Fig. \ref{fig:construction}).
The assembly utilizes a primary concave mirror that focuses light through a central aperture. A secondary planar fold mirror is positioned to redirect the optical path, significantly reducing the system’s longitudinal footprint and simplifying mechanical alignment. In our prototype, the fold mirror is secured to a planar glass substrate, providing a stable reference for precise, repeatable capture.
The sensor is interfaced with a motorized linear translation stage aligned with the optical axis. By translating the sensor, we acquire a focal stack that samples the non-planar image surface. This axial sweep is a fundamental requirement of the system; because simple reflective objectives exhibit pronounced Petzval curvature, on-axis and off-axis regions reach their respective focal planes at different depths.
Capturing the full focal stack ensures that every point across the wide field of view is sampled at its optimal focus. All data is recorded using a Sony IMX990 VIS-SWIR sensor, enabling broadband intensity measurements across the 400–1700nm spectrum.

\subsection{Focus Stacking}
Due to Petzval curvature, a simple mirror objective cannot form a sharp image across the entire field of view (FoV) on a flat sensor (see Fig.~\ref{fig:construction}(a)). One straightforward approach is to capture a dense focal stack that fully samples the volume containing the curved image surface, a technique previously explored by Matsunaga et al.~\cite{matsunaga2015field}.
This approach is often impractical for real-world deployment, as it typically requires 30 or more exposures to adequately cover the Petzval surface. Furthermore, Matsunaga et al. relied on spatially invariant deconvolution to address residual blur. This remains insufficient for high-aperture reflective systems, where the "sharpest" points on the Petzval surface still suffer from significant spatially varying aberrations, such as coma and astigmatism, that necessitate a more sophisticated, non-uniform PSF model.

\paragraph{Sparse focal stack capture.} We instead capture a sparse focal stack consisting of only $N$ images (typically 2-4), to span the depth of the Petzval curvature.
We demonstrate that performing joint multi-image deconvolution on this sparse set yields image quality comparable to, or exceeding, that of a dense stack while significantly reducing acquisition time and data overhead.

By mounting the mirror objective on a motorized translation stage, we can sample the focal volume at arbitrary axial positions.
We first move the translation stage that enables the sharpest view in the center of the FoV. Let this position be $z=z_0$, and let the corresponding image be $I_{z_0}(\bfp) \in \mathbb{R}^{H\times W}$, where H and W are number of rows and columns in the image respectively and $\bfp$ is the 2D coordinate in the image.
We then capture a total of $N$ images, each separated by a constant axial step $\Delta z$. 
The images in the focal stack are hence, $\left\{ I_{z_0 +k\Delta z}(\bfp)\right\}_{k=0}^{N-1}$.
$N$ typically depends on the curvature of the field. In practice, we found  that $N=3$ is sufficient to ensure that every point $\bfp$ in the FoV is sampled near its optimal focus in at least one measurement.
Consequently, at each spatial coordinate, the stack provides $N$ distinct measurements of the scene, allowing us to treat the recovery of a single sharp image, $I_0(\mathbf{p})$, as a well-posed multi-image inverse problem. To solve this, we first define a computationally tractable forward operator capable of modeling the severe, spatially varying off-axis aberrations across the stack.

\subsection{\methodname: Physics-inspired PSF modeling}
The off-axis aberrations at each focal stack position stem from astigmatism, coma, and often spherical aberrations, which are often modeled as Seidel aberrations~\cite{BornWolf1999}.
Approaches such as patch-wise PSF modeling~\cite{kee2011modeling} or Eigen decomposition of PSF~\cite{yeo2025eigencwd} are ill suited to model such extreme off-axis aberrations spanning several tens of pixels.
While microscopic or ideal systems are often modeled with radial or circular symmetry~\cite{gibson1991experimental}, real-world reflective assemblies frequently suffer from mechanical misalignment during mounting. This breaks the theoretical symmetry of the system, necessitating a flexible, learned model that can characterize arbitrary, non-symmetric PSF variations across the sensor.

\begin{figure}[!tt]
    \centering
    \includegraphics[width=\textwidth]{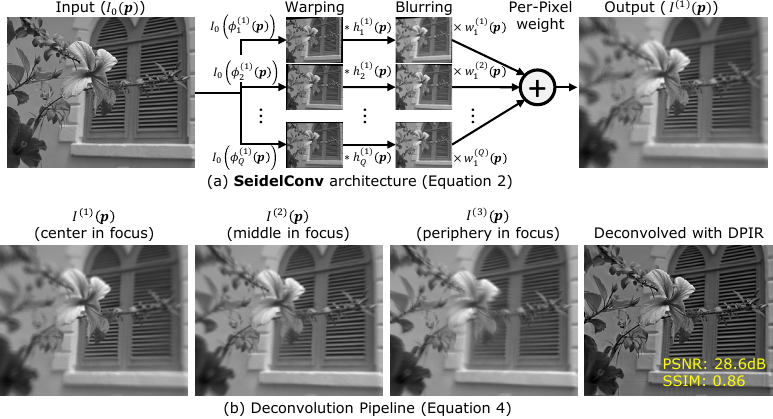}
    \caption{\textbf{\methodname architecture.} (a) shows \methodname architecture described in eq.~\eqref{eq:forward} for learning the forward operator for a single wavelength range. The input image is warped by an affine  (inspired by Seidel aberrations), and each warped image is blurred and weighed in a per-pixel manner and summed to get the output. (b) shows three images captured with our optical setup for the green wavelengths, and the deconvolved result by combining \methodname with a plug-and-play prior~\cite{zhang2021plug}, described in eq.~\eqref{eq:inverse_problem}. Each of the three focus settings captures a near sharp image in various regions. Our multi-focal deconvolution algorithm generates a single sharp image that is focused over the full FoV.}
    \label{fig:deconv}
\end{figure}

Inspired by the classical Seidel aberrations~\cite{BornWolf1999} of optical systems, and recent works in atmospheric turbulence modeling~\cite{chimitt2020simulating}, we propose a novel spatially varying PSF model designed to characterize the severe aberrations inherent in reflective optics.
We formulate each measurement in the focal stack as:
\begin{equation}
I^{(k)}(\bfp) = I_{z_0 + k\Delta z}(\bfp) = \mathcal{A}^{(k)}(I_0(\bfp)) + \eta^{(k)},
\end{equation}
where $\eta^{(k)}$ accounts for noise and model mismatch, and $\mathcal{A}^{(k)}$ is a spatially varying blur operator.
Our operator, \methodname, parametrizes the local PSF as a mixture of a small set of basis kernels with spatially varying weights and bounded affine deformation.
The bounded affine transforms are similar in spirit to the Seidel aberrations that are expressed as function of spatial (polar) coordinates. However, \methodname is sufficiently flexible to account for practical systems where the optics are potentially misaligned.
The \methodname forward operator for the $k^\text{th}$ image is,
\begin{align}
    I^{(k)}(\bfp) = \sum_{q=1}^{Q} \underbrace{w_q^{(k)}(\bfp)}_\text{per-pixel weight} \underbrace{I_0(\phi_q^{(k)}(\bfp))}_\text{warping} \ast \underbrace{h_q^{(k)}(\bfp)}_\text{blurring}, 
    \label{eq:forward}
\end{align}
where $w_q^{(k)}$ is the per-pixel weight, $\phi_q^{(k)}$ is a coordinate warping operator (in our case, an affine function), and $h_q^{(k)}(\bfp)$ is the blur kernel, all for the $k^{th}$ image (See Fig.~\ref{fig:deconv} (a)).
We model the warping operator as an affine function,
\begin{align}
    \phi_q^{(k)}(\bfp) = R_q^{(k)}\bfp + \textbf{t}_q^{(k)}.
\end{align}
Note that in the absence of affine warping, \methodname is similar to CoordGate~\cite{howard2024coordgate} or a learning variant of EigenCWD~\cite{yeo2025eigencwd}. However, the inclusion of the warping operation provides a critical geometric degree of freedom, allowing the model to precisely characterize the stretching and rotation of PSFs (e.g., coma and astigmatism) as they move off-axis. As demonstrated in our results, this makes \methodname considerably more effective at modeling the extreme aberrations typical of high-aperture mirror optics.

\paragraph{Multi-Image Deconvolution.}
Assuming the parameters of the PSF model are learned, we solve a multi image problem to estimate the latent sharp image $I_0(\bfp)$ by solving a simple optimization problem,
%

\begin{equation}
\min_{I_0(\bfp)}\; \sum_{k=1}^{N} \left\| \mathcal{A}^{(k)}(I_0(\bfp)) - I^{(k)}(\bfp) \right\|_2^2 + \lambda \, \mathcal{R}(I_0(\bfp)),
\label{eq:inverse_problem}
\end{equation}
where $\mathcal{R}(\cdot)$ is an image prior, which could be a simple total variational loss, a deep image prior, or, as in our case, a plug-and-play denoiser. 
This multi-image deconvolution enables considerably higher quality results compared to converting a focal stack into a single sharp image, or averaging the focal stack into a single blurred image with spatially invariant kernel (See Fig.~\ref{fig:deconv}(b)).

\begin{figure}[!tt]
    \centering
    \includegraphics[width=\linewidth]{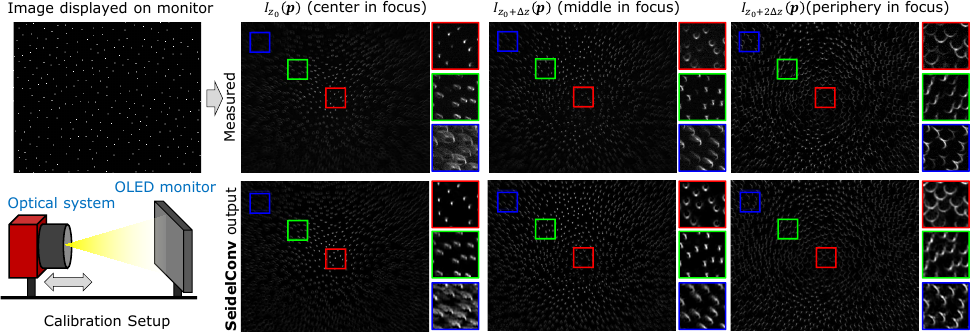}
    \caption{\textbf{Accuracy of PSF calibration.} The first column shows the calibration setup, where we displayed an image with sparse and random dots on an OLED monitor. The next three columns shows a random set of dots for three focus settings, (top row) measured by the optical system and (bottom row) output by \methodname. The model accurate captures the highly varying PSF with an error of $4\times10^{-3}$, enabling accurate deconvolution across the full field of view.}
    \label{fig:model}
\end{figure}

\paragraph{\methodname calibration.}
To calibrate the parameters $\theta_k$ (affine matrices, blur kernels, and per-pixel weights) of $A^{(k)}$, we leverage a monitor-based calibration approach~\cite{lin2025learning}.
Specifically, we display a random set of dots on a monitor placed a fixed distance away from the optical setup.
Then, we measure focal stack with relevant $z_0$ such that the center is in focus, along with a total of $N$ images placed $\Delta z$ apart (See Fig.~\ref{fig:model}). 
We then solve a stochastic gradient descent problem with the following optimization function,
\begin{equation}
\min_{\{\theta_k\}_{k=0}^{N-1}}\; \sum_{l=1}^{L} \sum_{k=1}^{N} \left\| \mathcal{A}^{(k)}_{\theta_k}(I_0^i(\bfp)) - I^{(k),i} \right\|_2^2
\;+\; \lambda_{\mathrm{kern}} \sum_{k=1}^{K} \|h_k\|_1,
\label{eq:forward_train}
\end{equation}
for $L$ calibration images. We regularize the blur kernels by imposing a $\ell_1$ norm to ensure sparsity. In our experiments, we used $L=10$ images for training the parameters, for $N=3$ focal stack images. Further details about calibration, including offset and vignetting calibration are available in the supplementary.
Note that after calibration, the subject can be at any arbitrary $z_0$, and not necessarily the same as the monitor distance. However, the spacing  between focal stack, $\Delta z$ is to be fixed.
Fig.~\ref{fig:model} shows the measurements for 50mm F/1 mirror at three focus settings and the output of the learned \methodname operator.

\paragraph{Generalization of PSF model to arbitrary wavelength range.} A unique advantage of mirror-based optics is that the PSF is nearly invariant to wavelength. This implies \methodname can be calibrated over any range of wavelengths and applied to any other range of wavelengths. We calibrated \methodname using broadband visible light (400–700nm) and successfully deconvolved images across the entire VIS-SWIR spectrum (400–1800nm) without fine-tuning.
This flexible capability is impractical with glass optics, where material dispersion necessitates unique focus settings or separate PSF models for each spectral band. Our approach bypasses these bottlenecks, enabling seamless cross-spectral imaging with a single hardware calibration.

%% file: results.tex

We now show experimental validation of our system using the prototype shown in Fig.~\ref{fig:construction}. All the results in this section are real captures. Panchromatic (grayscale) images were captured without any optical filter. For multispectral experiments, we utilized a filter wheel equipped with 2-inch spectral filters, including Blue, Green, Red, NIR, and SWIR, positioned in front of the objective.
The initial axial position, $z_0$, was optimized per-scene to ensure the center of the field of view (FoV) was at peak focus. The translation step size, $\Delta z$, was held constant at $200\,\mu\text{m}$ for all experiments with 50mm objective, and $100\mu\text{m}$ for all experiments with 100mm objective.
For the \methodname architecture, we utilized a mixture of $Q=31$ basis components, each consisting of an affine transformation and an $11 \times 11$ blur kernel. The model was implemented in PyTorch and trained on a calibration set of 10 images using a learning rate of $5 \times 10^{-3}$.
For the reconstruction phase, we employed the grayscale version of DPIR~\cite{zhang2021plug} as our plug-and-play prior. Notably, the same model was applied across all five spectral bands individually without any per-band fine-tuning, demonstrating the robust achromaticity of our reflective design.
Additional details are available in supplementary.


\paragraph{Deconvolution performance.}
Fig.~\ref{fig:deconv} shows three captured images, and the resultant deconvolved image with \methodname. The three images exhibit sharpness in various regions of the FoV, as a result of petzval curvature. Effectively, the focal stack images are intersecting the petzval surface at various points.
\methodname is able to recover a sharp image across the full field of view. 
Additional evaluations on the effect of parameters are in supplementary.
%
%

\begin{figure}[!tt]
    \centering
    \includegraphics[width=\textwidth]{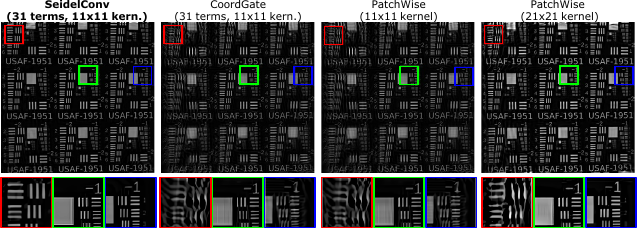}
    
    \vspace{1em}
    \resizebox{0.9\columnwidth}{!}{%
    \setlength{\tabcolsep}{4pt}
    \begin{tabularx}{1.1\linewidth}{@{}llcc@{}}
    \toprule
    \multirow{2}{*}{\# Focal stack images} & \multirow{2}{*}{PSF model} & \multicolumn{2}{c}{Metrics} \\
     & & On-Axis (P/S) & Off-Axis (P/S) \\
    \midrule
    \multirow{4}{*}{3 (\textbf{Computational mirrors})}
    & \textbf{\methodname} (11$\times$11 kern.)    & 27.6/\textbf{0.84} & \textbf{27.1}/\textbf{0.74} \\
    & CoordGate         & 27.6/0.80       & 24.8/0.63 \\
    & Patch-wise (11$\times$11 kern.) & 27.6/0.80                   & 24.8/0.63 \\
    & Patch-wise (21$\times$21 kern.) & \textbf{27.9}/0.80       & {21.6}/{0.65} \\
    \midrule
    \midrule
    $20\rightarrow1$~\cite{yokoya2015extended}\\(Focal stack average) & \textbf{\methodname} & 27.3/0.74 & 23.1/0.64 \\
    \midrule
    $20\rightarrow1$ ~\cite{matsunaga2015field} \\(Image on Petzval surface) & \textbf{\methodname} & 27.6/0.80 & 23.9/0.67 \\
    \midrule
    1~\cite{heide2013high} & \textbf{\methodname} & 25.6/0.66 & 20.9/0.49 \\
    \bottomrule
    \end{tabularx}
    }%
    \caption{\textbf{Comparison against other PSF models.} The figure above shows deconvolution results with various PSF models including \methodname, CoordGate~\cite{howard2024coordgate} that does not include warping, and two variants of patch-wise convolution with two kernel sizes. The table reports PSNR and SSIM evaluated on-axis (green box) and off-axis (red and blue box). All baselines perform similarly on axis, where the PSFs are compact. However, for off-axis, \methodname outperforms other models, demonstrating its importance for computational mirrors. The next three rows compare prior hardware-based approaches with \methodname PSF model. We observe that a multi-image deconvolution outperforms single-image, focal stack sum, or imaging on the Petzval surface.}
    \label{fig:psfmodels}
\end{figure}

\paragraph{Comparison against other PSF models.} To evaluate the efficacy of \methodname, we compared its performance against several state-of-the-art PSF models while maintaining identical deconvolution parameters across all tests. We curated a validation dataset by displaying 20 images from the Kodak dataset~\cite{kodak1999} on a radiometrically calibrated OLED display. To ensure strict radiometric accuracy, we utilized only the green channel. Performance was quantified using Peak Signal-to-Noise Ratio (PSNR) and Structural Similarity Index (SSIM). 
As shown in the USAF test target comparison (Fig.~\ref{fig:psfmodels}, top row), the importance of coordinate warping is clear. CoordGate model fails to compensate for extreme off-axis aberrations. While patch-wise convolution with a $21 \times 21$ kernel provided the second-best results, it remained qualitatively inferior to \methodname.
The table in Fig.~\ref{fig:psfmodels} shows metrics for the baselines including various PSF models for multi-image deconvolution with three images, deconvolution by summing up the focal stack~\cite{yokoya2015extended}, deconvolution of the Petzval curvature image~\cite{matsunaga2015field}, and a single image deconvolution with one image in the focal stack~\cite{heide2013high}.
We evaluated both on-axis metrics (center half) and off-axis metrics (everything other than center half).
While on-axis performance is comparable across most methods, \methodname significantly outperforms all baselines (more than 2dB higher) on off-axis. This highlights that affine warping is the key enabling factor for correcting the severe geometric distortions in wide-field reflective optics.

\begin{figure}[!tt]
    \centering
    \includegraphics[width=\linewidth]{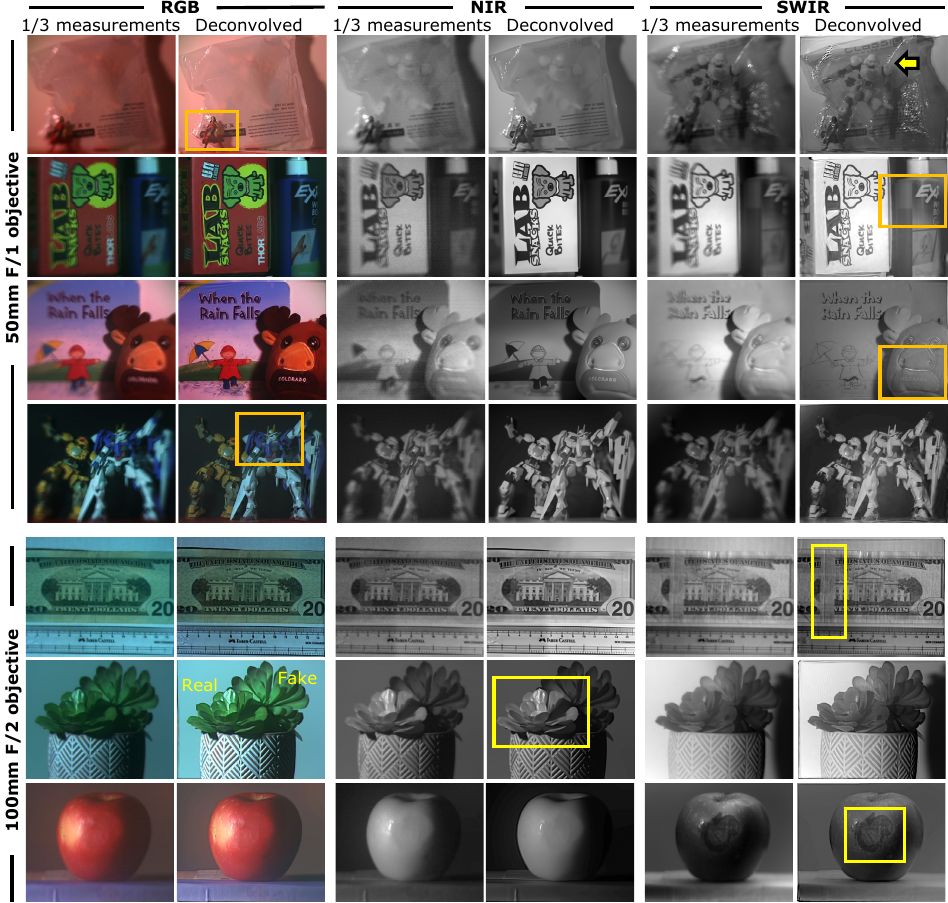}
    \caption{\textbf{High-resolution VIS-SWIR multispectral imaging.} The figure above shows raw images and reconstructions with the mirror optic in RGB, NIR, and SWIR wavelengths for 50mm F/1 objective, and 100mm F/2 objective. Data was captured with the same focus settings, and five different spectral filters (red, green, blue, NIR, SWIR). Individual images were then deconvolved with 3-image focal stack. Each set of wavelengths captures complementary information. RGB captures texture, NIR captures material and foliate information (real vs. fake plant), and SWIR is excellent for imaging through thin occluders (plastic bags and bottles) and for forensics (20 dollar note). Our computational correction enables sharp images independent of choice of wavelength with only three images per wavelength, requiring no brittle wavelength-specific calibration, or wavelength-specific refocusing.}
    \label{fig:ms_50mm}
\end{figure}

\paragraph{Multispectral imaging results with 50mm objective.}
Fig.~\ref{fig:ms_50mm} (top rows) shows various scenes captured in the lab with our optical setup. Row 1 shows an opaque bag with a robot inside that is only visible in SWIR image. Row 2 shows a bottle of liquid that is once again only visible in SWIR. Row 3 shows a book and a plush toy. The plush toy loses all color in SWIR. Finally, row 3 shows two colorful robots with rich spatial texture.
Across the board, our 50mm F/1 objective enables sharp images from 400nm all the way to 1700nm with a single focal setting. \methodname ensures that the resultant images are sharp across the whole field of view. The advantages of such an optic are evident in the rich and diverse set of features captured in individual wavelengths.



\begin{figure}[!tt]
    \centering
    \includegraphics[width=0.9\linewidth]{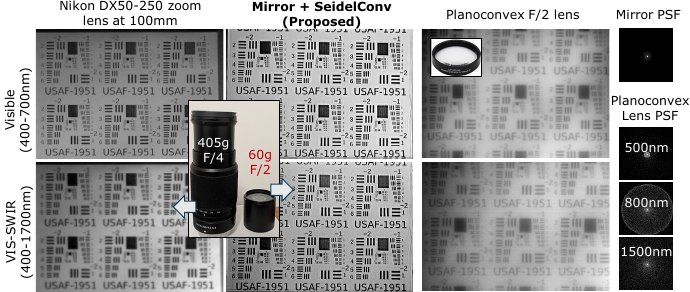}
    \caption{\textbf{Advantages of Mirrors.} We compare visible and broadband (400 - 1700nm) images for a (first column) SLR objective, (second column) proposed computational mirrors, and (third column) a plano-convex lens, all at 100mm focal length. The scene consisted of a printed USAF test target illuminated by broadband light. SLR objectives  enable very sharp images in visible wavelengths (where they are optimized for), but are bulky and produce blur over broadband wavelengths. Plano-convex lenses enable better off-axis performance but even their visible performance is considerably poor. This primarily arises due to high variance in point spread function across the various wavelengths, as shown in the fourth column. Our computational mirrors are enabled by very light optics, and produce a sharp image across all wavelengths and full FoV.}
    \label{fig:baselines}
\end{figure}

\paragraph{Long range imaging.}
Fig.~\ref{fig:ms_50mm} (bottom rows) shows results with 100mm F/2 optics with other configuration similar to the 50mm optic. A separate set of forward operators were learned for the 100mm optic. Objects were placed 6-10 feet away to demonstrate material identification at long distances.
First row shows a genuine \$20 note, where the SWIR image shows a blank band to differentiate it from counterfeit notes. 
Second row shows a real (foreground) and artificial plant, that look nearly the same in RGB and SWIR wavelengths, but the real plant looks considerably brighter in NIR wavelengths. 
Third row shows an apple with a bruise, which is not visible in RGB or NIR, but distinctly visible in SWIR.
Though the blur is less severe than 50mm objective, it nevertheless results in loss of important details.
All these results were captured with the same focus setting ($z_0$), implying there was no wavelength-dependent refocusing.

\paragraph{Comparison against glass-based objectives.}
To show the advantages of mirror-based optics for broadband VIS-SWIR imaging, Fig.~\ref{fig:baselines} compares it against a mirrorless SLR 100mm zoom lens, and a plano-convex F/2 lens. Images captured by mirror optic were reconstructed with \methodname. We observe that Nikon lens produces sharpest image within the visible wavelengths, but has a blur when imaged over the full VIS-SWIR Wavelengths. Plano-convex in contrast has blurry image across any wavelength range owing to its severe chromatic aberration. In contrast, our mirror based optic produces sharp images independent of range.
Additionally, our mirror-based optic weighs only 60g and has a 50mm  length, whereas a 100mm Nikon lens weights more than 400g and is thrice as long.

\begin{figure}[!tt]
    \centering
    \includegraphics[width=\linewidth]{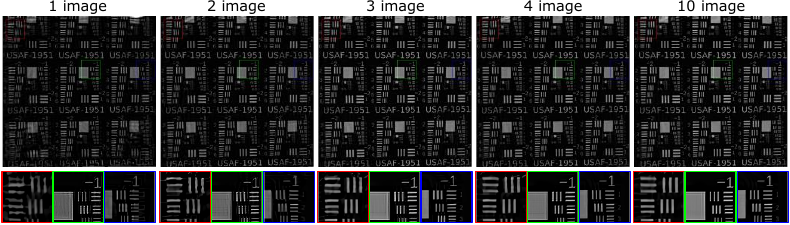}
    \caption{\textbf{Effect of number of images in focal stack.} We show results for varying number of images in the focal stack for a 50mm, F/1 optic. As few as three images results in a dramatic improvement in off-axis resolution, compared to a single image. Additionally, performance plateaus after 3 images, making three a heuristically optimal number of images in this scenario.}
    \vspace{-2em}
    \label{fig:num_images}
\end{figure}

\paragraph{Number of images in the focal stack.} Fig.~\ref{fig:num_images} shows deconvolution on our 50mm mirror objective for varying number of images in the focal stack. We notice that the quality improves with increasing number of images, but saturates after 3 images, motivating most of the experiments in this paper. 
It is worth noting that the number of images depends on the field curvature. For the same sensor size, a 25mm objective will have twice the curvature of a 50mm objective, likely necessitating more images in the focal stack. On the other hand, a 100mm objective has lower curvature, implying even two images may suffice.
For longer focal lengths, we anticipate that even a single image can result in sharp images.

%% file: conclusion.tex
This work validates Computational Mirrors as a robust framework for high-aperture, ultra-broadband imaging. Using $f/1$ 50mm and $f/2$ 100mm prototypes, we demonstrated that a sparse focal stack of only three images is sufficient to reconstruct high-fidelity, all-in-focus imagery across a wide field of view. The core of this capability is \methodname, a physically motivated and computationally efficient PSF model that, when paired with a plug and play reconstruction framework, effectively mitigates the severe geometric aberrations of simple reflective optics. Our results across the VIS, NIR, and SWIR spectra highlight the primary advantage of this approach: because reflective optics are inherently achromatic, they require no wavelength-dependent refocusing or PSF recalibration. By bridging the gap between simple, lightweight mirrors and advanced VIS-SWIR sensors, our method facilitates high-performance material characterization and identification in a compact form factor that was previously unattainable with refractive designs.




%% file: supp/hardware.tex
\subsection{Camera and translation stage details}
\paragraph{Camera.} We equipped our mirror objective with a Touptek SenSWIR camera that can measure light intensity from 400 to 1700\,nm.
The sensor had a resolution of $1024\times 1280$. We binned the sensor by a $2\times2$ window which resulted in $512\times640$ images. This was primarily done to reduce computational load.
Our approach can be applied for full resolution as well, albeit slightly slower to recover.

\paragraph{Translation stage.} We used a Thorlabs Elliptec ELL17 translation stage that is fast (180\,mm/s), has a maximum translation of 28\,mm, and a position accuracy of $50\mu\,m$. However, we found that we could operate it with $20\mu\,m$ steps, as it had a position accuracy of $20\mu\,m$. This enabled us to capture a dense focal stack to evaluate prior works including Yokoya and Nayar~\cite{yokoya2015extended}, and Matsunaga and Nayar~\cite{matsunaga2015field}.
An alternative solution could be to mount the folding mirror alone on a stepper motor translation stage, which can be considerably faster, and lead to a more compact solution.

\subsection{Mirror objective details}
We bought off-the shelf concave mirrors from Edmund optics (EO \#43-470 for 50\,mm and \#43-841 for 100\,mm), both protected aluminum that can has high reflectivity between 400\,nm and 20$\mu\,m$, making it an ideal and low-cost (< \$100) option.
For the aperture in the center, we used a drill press equipped with a diamond tipped hole saw to ensure the glass did not crack. The drill press was run at 700 RPM while constantly spraying glass cleaning fluid to keep heat to a minimum.
A folding mirror (EO \#89-482) was glued to a transparent two-inch window (EO \#83-366) that transmits from 200 - 2000\,nm.
For transmission beyond these wavelengths, a better solution would be a metal trellis with spider vanes, enabling all wavelengths.
Finally, all the optical elements were encased in a Thorlabs SM-2 barrel and secured in 60\,mm cage plates.

\subsection{PSF calibration}
We now detail the process of PSF calibration and the data capture pipeline.

\begin{figure}[!tt]
    \centering
    \includegraphics[width=\columnwidth]{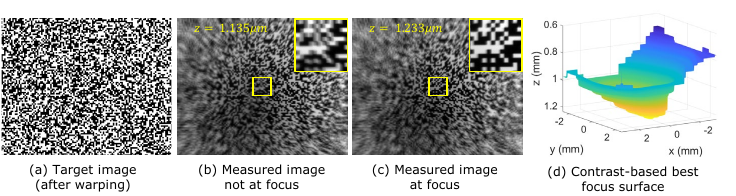}
    \caption{\textbf{Estimating best focus properties.} We estimated best focus distance and the range of focal sweep for optimal results. (a) We capture a dense focal stack of a random binary target at 20$\mu\,m$ steps over a 2\,mm range, and then use a contrast operator to estimate the best focus distance (b, c). (d) We also estimate a ``best focal surface to estimate the range of focal sweep. In our case, $400\mu\,m$ was optimal for 50\,mm objective, and $200\mu\,m$ was optimal for 100\,mm objective. }
    \label{fig:focus}
\end{figure}
\subsubsection{Best focus settings}
We first estimate the optical center of the image, best focus setting, and the range of z values over which we can ensure optimal reconstruction quality.
To do this, we capture a random binary pattern (see Fig.~\ref{fig:focus}) that enables estimation of sharpness over the full field of view, for each focal setting.
We then sweep over a 2\,mm range in 20$\mu\,m$ steps to get a dense focal stack.
We used this dense focal stack to first get the focus distance that enables the sharpest image in the center (b, c in the figure).
We used this focus setting for calibrating monitor-to-sensor homography in the next step.
We note that this optimal focus setting changes from scene to scene, depending on where the objects are placed. However, the calibrated \methodname works for all object distances.
Additionally, we also estimated a ``best focus surface" using a local contrast operator, as shown in subfigure (d).
We used this best focus surface to estimate the range of z values that are in focus.
Across the field of view, we found that $400\mu\,m$ was an optimal distance for the 50\,mm objective, and $200\mu\,m$ was optimal for 100\,mm objective.
It is worth noting that the focus surface is neither radially sy\,mmetric, nor the theoretical spherical surface. This is likely due to mechanical misalig\,nments.
Invariably, approaches such as SiedelNet~\cite{kohli2025ring} that rely on radial sy\,mmetry will not work.
However, \methodname does not need radial sy\,mmetry, and hence any misalig\,nments are well-modeled by our approach.

\subsubsection{Monitor-to-sensor calibration}
\begin{figure}[!tt]
    \centering
    \includegraphics[width=\columnwidth]{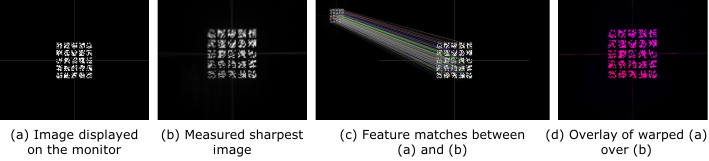}
    \caption{\textbf{Monitor-to-sensor calibration.} We estimate a homography transformation between monitor and sensor. (a) We display a cross-hair combined with a $25\times25$ ARuCo tile which has several unique spatial features. The captured image in (b) is then used to perform feature matching, as shown in (c). (d) shows the warped monitor image overlaid over the measured image in (b), showing a near-perfect match in the center of the field of view, where the image is the sharpest.}
    \label{fig:homography}
\end{figure}

We estimate a homography transformation between monitor and image sensor to perform PSF calibration, and for evaluating performance metrics over the kodak image dataset~\cite{kodak1999}. 
Figure~\ref{fig:homography} shows our pipeline for estimating the homography matrix.
Since the central part of the FoV is in focus, we display a $5\times5$ grid of ARuCo markers spanning the central third of the field of view, or less.
We then capture an image with the best focus distance estimated in the previous step.
These two images are used for feature matching, from which we estimate the homography.
This center-based homography estimation enabled precise calibration, as shown in subfigure (d), with an overlay of measured image, and warped monitor image.

\subsubsection{Vignetting and offset}
\begin{figure}[!tt]
    \centering
    \includegraphics[width=\columnwidth]{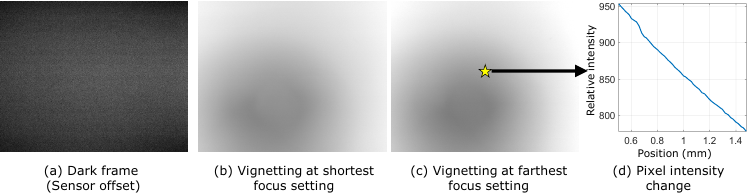}
    \caption{\textbf{Offset and vignetting.} Figure (a) shows the image sensor's dark frame, with fixed pattern noise, which we subtract from every measurement. (b) shows vignetting when the objective is closest to the sensor and (c) shows when the objective is farthest. While similar in relative intensity distribution, we see a change in intensity value, as shown in (d). We collected a one-time vignetting stack and used the appropriate focus setting for all our experiments.}
    \label{fig:radiometry}
\end{figure}

Doped InGaAs sensors tend to suffer from fixed pattern noise, particularly when measuring visible wavelengths, as seen in Fig.~\ref{fig:radiometry}.
We estimated this ``dark frame" by closing the camera and optics with a shutter and measuring the image.
Additionally, we measured vignetting due to the mirror optics, which can cause the center of the image to appear darker (see subfigure (b) and (c)). 
While the intensity distribution remains the same, the per-pixel intensity changes as a function of focus distance (subfigure (d)).
We therefore measure a dense set of vignetting frames, and use the appropriate frame at each focal setting.

\subsection{Multispectral capture}
\begin{figure}[!tt]
    \centering
    \includegraphics[width=\columnwidth]{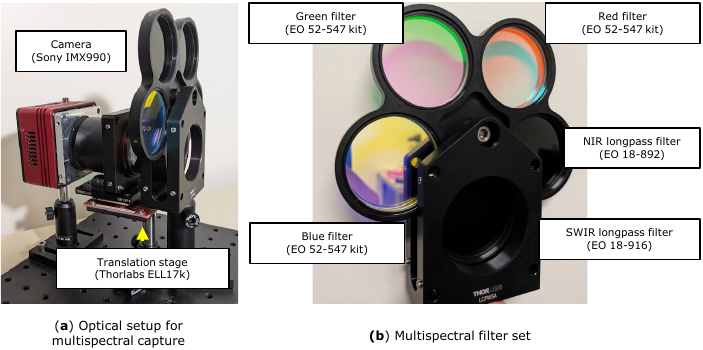}
    \caption{\textbf{Multispectral capture setup.} We used red, green, blue absorptive filters, and NIR and SWIR longpass filters. To obtain NIR bandpass images, we illuminated the scene with a compact fluorescent light, which had intensity only till 800\,nm.}
    \label{fig:ms_setup}
\end{figure}
Figure~\ref{fig:ms_setup} shows the setup for capturing multispectral data with our optical setup.
We used three absorptive color filters in combination with hot mirrors to reduce leakage from SWIR wavelengths.
Additionally, we used an NIR longpass filter and a SWIR longpass filter.
For capturing red, green, blue, and NIR (bandpass) images, we used a white compact fluorescent lamp. 
This resulted in a small drop in signal-to-noise ratio in the NIR measurements, but ensured a bandpass measurement with a longpass filter.
For SWIR images, we illuminated the scene with an unfiltered halogen lamp.

%% file: supp/results.tex
\subsection{Effect of Number of Focal Stack Images}

\begin{figure}[!tt]
    \centering
    \includegraphics[width=\linewidth]{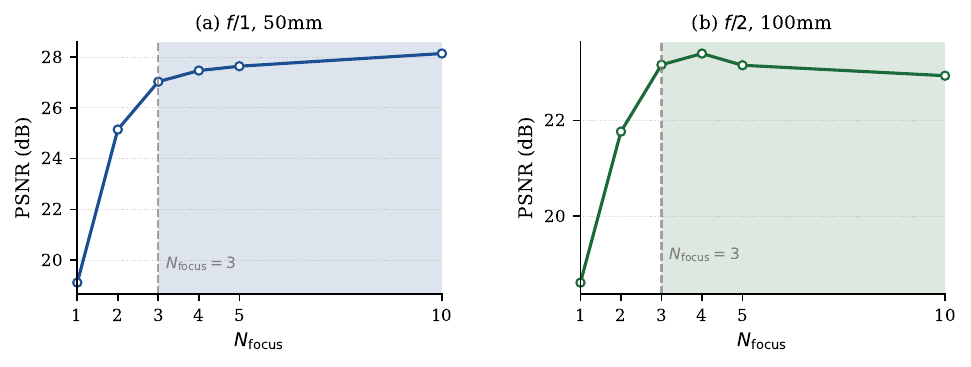}
    \caption{\textbf{Effect of focal stack size.} Full-image PSNR as a function of the number of focal stack images $N_{\text{focus}}$ for (a) the $f/1$ 50\,mm and (b) the $f/2$ 100\,mm mirror objectives. In both cases, performance improves sharply from a single image and saturates at $N_{\text{focus}}=3$ (dashed), beyond which the gains become marginal.}
\label{fig:nfocus}
\end{figure}

Fig.~\ref{fig:nfocus} shows full-image PSNR as a function of number of focal stack images ($N_{\text{focus}}$) for both mirror objectives. A single capture yields poor full-field quality, as no single sensor position can simultaneously cover the entire curved focus surface. Performance improves substantially with each additional image up to $N_{\text{focus}} = 3$, after which the gains become marginal. This confirms that a minimal focal stack of three images is sufficient to sample every field point at or near its focus, motivating our default choice of $N_{\text{focus}} = 3$.

\subsection{Baseline validation}
\begin{figure}[!tt]
    \centering
    \includegraphics[width=\linewidth]{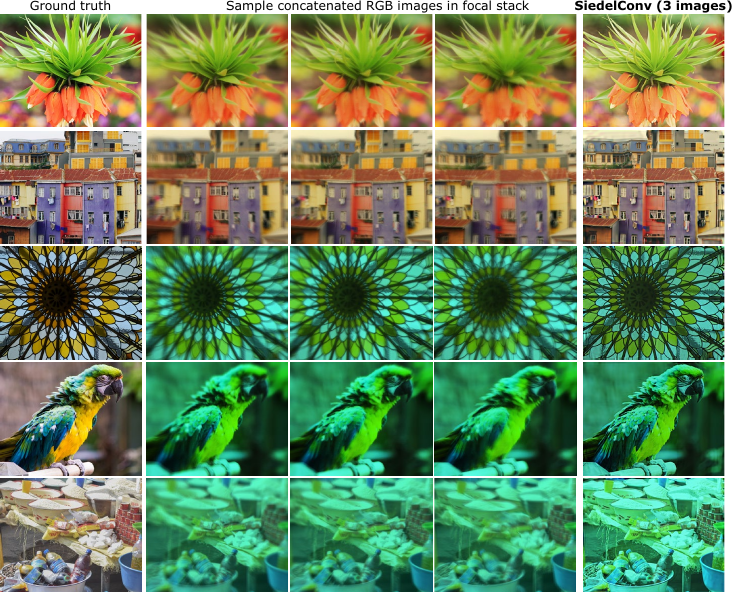}
    \caption{\textbf{Computational mirrors dataset.} To enable further research on computational mirrors, we will be releasing a 100-image dataset captured from the DIV2K validation dataset with our optical setup. The figure above shows ground truth, 2 sample images in the dense focal stack, and reconstruction with \methodname.}
    \label{fig:div2k}
\end{figure}
To encourage further research in mirror-based objectives, we will be releasing a 100-image dataset along with calibration files captured with our 50\,mm and 100\,mm objectives.
We collected red, green, and blue images by displaying individual channels of the validation set of DIV2k dataset~\cite{Agustsson_2017_CVPR_Workshops,Timofte_2017_CVPR_Workshops} consisting of 100 high-resolution color images.
We captured a dense focal stack containing 20 images for each channel, and performed offset and vignetting correction.
The individual channels were captured by displaying the corresponding channel on the monitor.
Figure~\ref{fig:div2k} shows ground truth, three of the focal stack images, and a sample reconstruction using \methodname.
We will make this dataset public along with \methodname code.

\subsection{Simple lens experiments}
\begin{figure}[!tt]
    \centering
    \includegraphics[width=\linewidth]{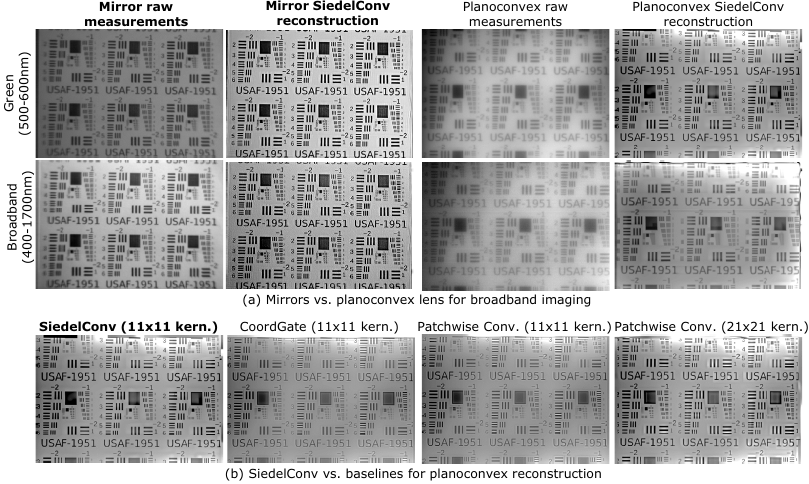}
    \caption{\textbf{Advantages of mirrors and \methodname.} (a) we compare mirror and planoconvex lens for broadband imaging. Mirrors are inherently broadband, implying that a model trained on green wavelengths can be used for any spectral range. Therefore, mirrors produce sharp results whether in green wavelengths, or full VIS-SWIR wavelengths. In contrast, planoconvex lenses, when trained on green wavelengths, enable sharp results on green wavelengths, but significantly poorer results for VIS-SWIR imaging. (b) We show that \methodname generalizes to any focusing element. We compare it in green wavelengths against CoordGate, and patchwise deconvolution for two kernel sizes, and observe that \methodname produces the best reconstruction accuracy.}
    \label{fig:pcx}
\end{figure}

Figure~\ref{fig:pcx} shows results with a 100\,mm planoconvex lens and compares it to our 100\,mm mirror objective. 
Similar to the mirror objective, we capture 3 images in the focal stack with a planoconvex lens.
Additionally, we capture two ranges of wavelengths, one over green-only, and another broadband over VIS-SWIR.
We observe that planoconvex data, when combined with \methodname, results in high-quality reconstruction in a small wavelength range (green).
However, when we apply the model to the broadband range of wavelengths, we observe that a planoconvex lens leads to severe blur, due to wavelength-dependent PSF.

Additionally, to demonstrate that \methodname generalizes beyond mirrors, we compare it against other baselines in subfigure (b). In this example, we deconvolved only the green wavelength.
We observe that \methodname produces significantly better reconstruction compared to other baselines, implying that \methodname can be used with nearly any focusing element, mirror or refractive.

\subsection{Effect of \methodname Hyperparameters}

We analyze the sensitivity of \methodname to its two primary hyperparameters, the number of affine terms $N_{\text{terms}}$, which controls the expressiveness of the spatially-varying warp field, and kernel size. All experiments are conducted on the $f/1$ 50\,mm mirror objective.

\begin{figure}[!tt]
    \centering
    \includegraphics[width=\linewidth]{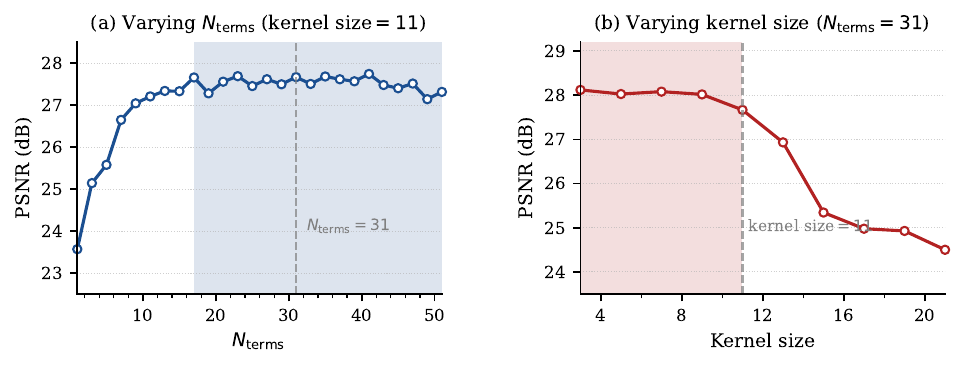}
    \caption{\textbf{Effect of \methodname hyperparameters.}
    (a) Full-image PSNR vs.\ number of affine terms $N_{\text{terms}}$
    (with kernel size $=11$): performance improves rapidly for small values
    and stabilizes beyond $N_{\text{terms}} \approx 17$. We adopt $N_{\text{terms}}=31$
    (dashed) as a conservative choice within the saturated regime.
    (b) Full-image PSNR vs.\ kernel size
    (with $N_{\text{terms}}=31$): performance is stable for small to moderate
    kernel sizes but degrades for kernel sizes $>11$ (dashed), as excessively
    large kernels introduce overfitting and reconstruction artifacts.
    We select kernel size $=11$, which provides a good trade-off between
    modeling capacity and stability, and was empirically found to be more
    robust for noisier near-infrared (NIR) measurements.}
    \label{fig:hyperparams}
\end{figure}

\noindent\textbf{Number of affine terms.}
Fig.~\ref{fig:hyperparams}(a) reports full-image PSNR as a function of $N_{\text{terms}}$ with kernel size $=11$. Performance improves substantially up to $N_{\text{terms}} \approx 17$, reflecting the need for multiple affine components to capture the spatially-varying off-axis distortions inherent to mirror optics. Beyond this point, gains saturate, indicating sufficient model 
capacity to represent the dominant aberration modes. We adopt $N_{\text{terms}} = 31$ as our default, balancing reconstruction quality and computational cost.

\noindent\textbf{Kernel size.}
Fig.~\ref{fig:hyperparams}(b) shows full-image PSNR as a function of the kernel size with $N_{\text{terms}}=31$. Performance is stable for small to moderate kernel sizes ($\leq 11$) but degrades for larger values. Increasing the kernel size provides additional modeling flexibility, but excessively large kernels tend to overfit local blur structure and introduce artifacts, particularly in off-axis regions where aberrations are strongest. While a larger kernel size can slightly improve on-axis PSNR, it reduces overall image quality. Based on this trade-off, we adopt kernel size $=11$ in all experiments. We also found that a kernel size of 11 is slightly more robust for near-infrared (NIR) wavelengths, where the captured images are noisier than in the visible range.

\begin{table}[t]
\centering
\caption{Full-image PSNR comparison of different reconstruction priors under the same calibrated multi-image spatially varying deconvolution model. Higher values indicate better reconstruction quality, with Plug-and-Play performing best.}
\label{tab:prior_effect}
\begin{tabular}{lc}
\toprule
Prior & PSNR (dB) \\
\midrule
Total Variation        & 23.4 \\
Deep Image Prior~\cite{ulyanov2018deep}       & 20.7 \\
INR-based prior~\cite{saragadam2023wire}  & 24.7 \\
PnP~\cite{zhang2021plug}       & \textbf{25.1} \\
\bottomrule
\end{tabular}
\end{table}

\begin{figure}[!tt]
    \centering
    \includegraphics[width=\linewidth]{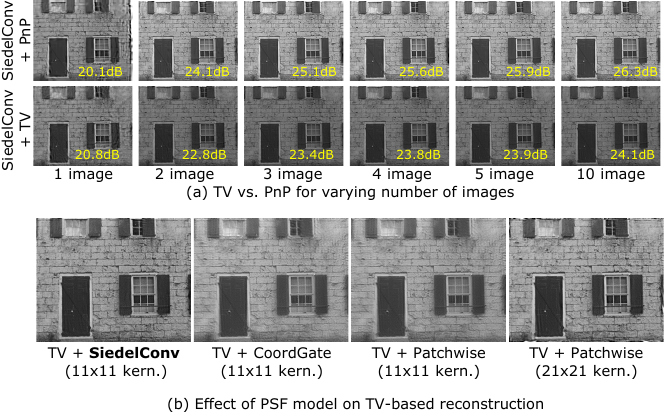}
    \caption{\textbf{Prior and \methodname.} (a) We compare the effect of prior on reconstruction. We observe that PnP enables better reconstruction than a simple total variation (TV) prior, informing the choice in the main paper. (b) We further evaluate the importance of \methodname. We used a TV-based reconstruction along with various PSF models. \methodname outperforms all other approaches even with a simple TV-based prior.}
    \label{fig:prior}
\end{figure}

\subsection{Effect of priors}
We compare several priors for the reconstruction stage while keeping the calibrated multi-image forward model fixed. Specifically, we evaluate Total Variation (TV), Deep Image Prior (DIP)~\cite{ulyanov2018deep}, an Implicit Neural Representation prior~\cite{saragadam2023wire}, and a Plug-and-Play (PnP) prior~\cite{zhang2021plug}.

As shown in Table~\ref{tab:prior_effect}, PnP achieves the highest full-image PSNR of 25.1\,dB, compared with 24.7\,dB for the INR-based prior, 23.4\,dB for TV, and 20.7\,dB for the Deep Image Prior. This result supports our choice of PnP as the main reconstruction approach, since it provides the strongest prior for this problem.

Figure~\ref{fig:prior}(a) compares PnP against TV-based reconstruction for varying number of images. We observe that PnP enables higher reconstruction quality with increasing number of images, particularly 2 and higher, where the quality improves dramatically over a single image-based reconstruction.
However, we notice that the success of the reconstruction primarily depends on the PSF model.
Figure~\ref{fig:prior}(b) compares a TV-based reconstruction with various PSF models.
We notice that \methodname has better reconstruction than baselines, with sharper features and lower ringing artifacts.

\subsection{Effect of PnP Solver Hyperparameters}
\begin{figure}[!tt]
    \centering
    \includegraphics[width=\linewidth]{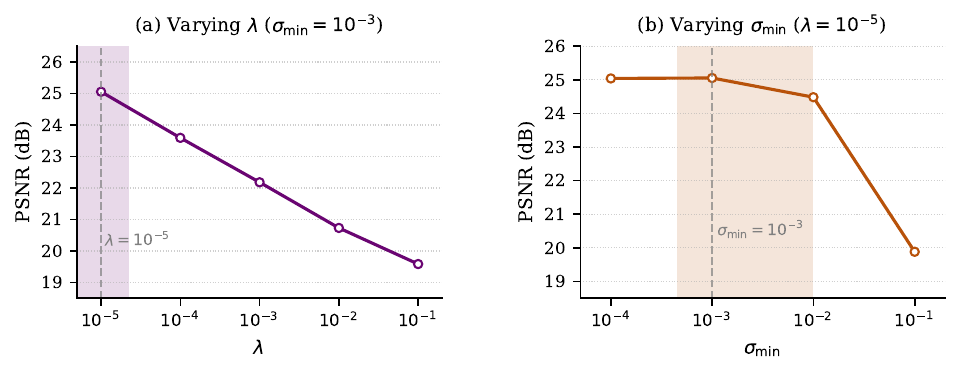}
    \caption{\textbf{Effect of PnP solver hyperparameters.}
    (a) Full-image PSNR vs.\ regularization weight $\lambda$
    ($\sigma_{\min}=10^{-3}$): reconstruction quality degrades monotonically
    as $\lambda$ increases, reflecting excessive penalization of the data term.
    Performance converges for small $\lambda$, and we adopt $\lambda=10^{-5}$
    (dashed) as a conservative choice in this stable regime.
    (b) Full-image PSNR vs.\ minimum noise level $\sigma_{\min}$
    ($\lambda=10^{-5}$): performance is stable for $\sigma_{\min} \in
    [10^{-4}, 10^{-3}]$ but drops sharply at larger values, where the
    denoiser is invoked too aggressively and over-smooths the solution.
    We adopt $\sigma_{\min}=10^{-3}$ as our default (dashed).}
    \label{fig:pnp}
\end{figure}

Fig.~\ref{fig:pnp} evaluates the two primary hyperparameters of the
Plug-and-Play solver: the regularization weight $\lambda$ and the minimum
denoiser noise level $\sigma_{\min}$.

\noindent\textbf{Regularization weight.}
Fig.~\ref{fig:pnp}(a) examines $\lambda$ with $\sigma_{\min}=10^{-3}$ fixed.
PSNR degrades monotonically as $\lambda$ increases, reflecting an
over-regularized solution that suppresses fine detail in favor of excessive
smoothness. We therefore adopt a small $\lambda=10^{-5}$, which provides
sufficient regularization without over-penalizing the data term.

\noindent\textbf{Minimum noise level.}
Fig.~\ref{fig:pnp}(b) examines $\sigma_{\min}$ with $\lambda=10^{-5}$ fixed.
Performance is stable for small values ($\sigma_{\min} \leq 10^{-3}$) but
drops sharply at $\sigma_{\min}=10^{-1}$, where invoking the denoiser at a very aggressive noise level causes it to over-smooth the solution and
discard fine structure. We adopt $\sigma_{\min}=10^{-3}$ as our default,
which keeps the denoiser in a well-behaved regime throughout the optimization.

\subsection{MTF analysis}
\begin{figure}[!tt]
    \centering
    \includegraphics[width=\linewidth]{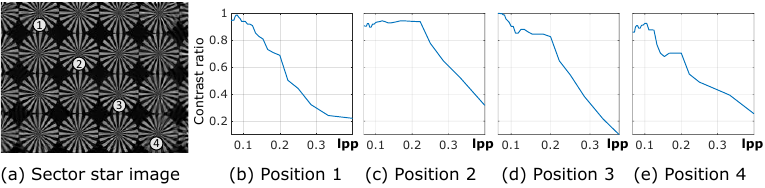}
    \caption{\textbf{Wide field-of-view imaging.} Here we show that we achieve an MTF30 (contrast ratio of 30\%) for a line pair of 0.3 or higher across the whole sensor, emphasizing the full usable field of view after computational reconstruction with a 3-image focal stack.}
    \label{fig:mtf}
\end{figure}

Figure~\ref{fig:mtf} shows MTF analysis across the full FoV with the 50\,mm objective.
We imaged a grid of sector star targets displayed on a monitor, and then estimated sharp images with a focal stack of three images, \methodname, and plug-and-play prior.
We then estimated the contrast over circles of varying radius from the center of each sector star.
The contrast was estimated as $(I_\text{max} - I_\text{min})/(I_\text{max} + I_\text{min})$, over each circle.
Across the full field of view, we observe an MTF30 (30\% or better contrast ratio) of 0.3 lines per pixels (0.5 is the maximum), emphasizing the efficacy of our approach.

